\documentclass[twocolumn,showpacs,prb,aps,showpacs]{revtex4}
\usepackage{dcolumn}
\usepackage{bm}
\usepackage[dvips]{graphics}
\begin{document}
\title{Superfluidity or supersolidity as a Consequence of Off-diagonal Long-range Order}
\author{Yu Shi} \affiliation{Center for Advanced Study,
Tsinghua University, Beijing 100084, China}

[Published in Phys. Rev. B {\bf 72}, 014533 (2005)]

\begin{abstract}
We present a general derivation of Hess-Fairbank effect or
nonclassical rotational inertial (NCRI), i.e. the refusal to
rotate with its container, as well as the quantization of angular
momentum, as  consequences of off-diagonal long-range order
(ODLRO) in an interacting Bose system. Afterwards, the path
integral formulation of superfluid density is rederived without
ignoring the centrifugal potential. Finally and in particular, for
a class of variational wavefunctions used for solid helium,
treating the constraint of single-valuedness boundary condition
carefully, we show that there is no ODLRO and, especially,
demonstrate explicitly that NCRI cannot be possessed in absence of
defects, even though there exist zero-point motion and exchange
effect.
\end{abstract}

 \maketitle

\section{Introduction}

It was first suggested by London that the ability of liquid $^4$He
II to flow through narrow capillaries without apparent friction is
a consequence of Bose-Einstein condensation (BEC).~\cite{london}
The concept of BEC was later generalized by Penrose and Onsager to
be applicable to interacting particles.~\cite{penrose0,penrose} It
was further generalized and systematically investigated by Yang,
as the notion of off-diagonal long-range order
(ODLRO).~\cite{yang} Now it is known that the no-friction behavior
in narrow capillaries is only one of several phenomena of
superfluidity.~\cite{leggett0} As elaborated by
Leggett,~\cite{leggett1} the most basic manifestation of
superfluidity is the Hess-Fairbank effect,~\cite{hess} which was
also called ``nonclassical rotational inertial'' (NCRI) by
Leggett.~\cite{leggett0} This refers to the refusal of the system
to rotate with its container, when its angular velocity is
sufficiently low. It is the counterpart of the Meissner effect of
superconductivity. Furthermore, the quantization of angular
momentum of the superfluid in the rotating container is the
counterpart of the magnetic flux quantization in a superconductor.

In the case of superconductivity, the demonstration of Meissner
effect and  the magnetic flux quantization, as consequences of
ODLRO,  was made by Yang,~\cite{yang} and by Sewell and Nieh {\it
et al.} in a more recent alternative approach.~\cite{sewell1}
Bloch discussed the relation between superconducting persistent
current and ODLRO.~\cite{bloch} In the case of superfluidity, Kohn
and Sherrington derived the Hess-Fairbank effect as a consequence
of ODLRO by using a sophisticated hierarchy of equations of
thermal Green functions.~\cite{ks} For a noninteracting Bose gas
and a Gross-Pitaevskii system, Leggett made a clear-cut
demonstration of Hess-Fairbank effect and quantization of angular
momentum as consequences of BEC.~\cite{leggett2}  Earlier, in an
extremely thorough and insightful discussion,~\cite{leggett0}
Leggett pointed out that a sufficient condition of superfluidity
is a certain topological connectedness property of the many-body
wavefunction, and that at least for zero temperature, ODLRO gives
rise to this connectivity and thus superfluidity, but for a finite
temperature, whether ODLRO is sufficient for superfluidity in
general is not conclusive.

Moreover, Leggett established, from the point of view of
connectivity of wavefunction, that BEC and NCRI behavior can  in
principle also be exhibited by a solid.~\cite{leggett3} Recently,
Kim and Chan clearly observed NCRI-like behavior in bulk solid
$^4$He in an annulus channel,~\cite{kim} shortly after an earlier
such observation in solid $^4$He confined in porous Vycor
glass.~\cite{kim1}  But a consensus on its origin is yet to be
reached.~\cite{leggetts,ceperley,prokofev,others} The earliest
predictions on supersolidity, i.e. superfluid behavior in a solid,
were based on BEC of defect states.~\cite{andreev,chester} But the
concentration of zero-point vacancies is less than $10^{-6}$
according to the experimental results.~\cite{meisel} Thus an
important question is whether it is possible for a pure
commensurate sample of solid $^4$He, i.e. without vacancies or
interstitials, become a supersolid. Negative answers were given
recently in a path integral Monte Carlo calculation of exchange
frequencies in bulk hcp $^4$He,~\cite{ceperley} and in a general
argument about superfluidity density.~\cite{prokofev}

Thus from  both the fundamental point of view and the  perspective
of understanding supersolid behavior, it appears still interesting
to make a general derivation of the Hess-Fairbank effect and
quantization of angular momentum as clear consequences of ODLRO
for an interacting Bose system in a rotating container. In this
paper, we first make such a derivation. Afterwards, for a reason
explained below, we rederive the superfluid density in the path
integral formulation,~\cite{pollock} which is the very basis of
the analyses of solid $^4$He in Ref.~\cite{ceperley} and
Ref.~\cite{prokofev}. Finally, we consider the trial wavefunctions
ever used in variational calculations for solid helium, including
the Hartree wavefunction, the Hatree-Fock wavefunction and the
Nosanow-Jastrow wavefunction. It is shown that there is no ODLRO
or BEC in these wavefunctions.  Moreover, by examining the
dependence of free energy on the rotation velocity of the
container, we explicitly demonstrate that a commensurate solid
described by such wavefunctions cannot possess NCRI, in absence of
vacancies or interstitials, even if there exist zero-point motion
and the exchange effect.

Note that the non-superfluidity of the Hartree-Fock wavefunction
made up of localized single particle wavepackets has been
discussed by Leggett from the point of view of disconnectivity of
the wavefunction long ago.~\cite{leggett0} Our approach provides
an explicit construction of the rotating wavefunction under the
constraint of the ``single-valuedness boundary condition'' (SVBC)
as called by Leggett~\cite{leggett0}, while keeping the energy the
same as that in the static case. To do this, adjustment on the
wavefunction needs to be made in the exponentially vanishing
regions, indeed as argued by Leggett.

The organization of the paper can be clearly seen in the section
titles.

\section{Hamiltonians and Free energies in the two reference frames}

As usual, consider a Bose system in a container rotating with a
angular velocity $\bm{\omega}$. Thermodynamic equilibrium is
determined by the minimization of the free energy in the
co-rotating frame of reference, in which the wall of the container
is at rest. In this frame, the Hamiltonian is
\begin{equation}
\begin{array}{ll}
H = & \displaystyle\sum_{j}
[\frac{(\mathbf{p}_j-m\bm{\omega}\times\mathbf{r}_j)^2}{2m}
-\frac{1}{2}m(\bm{\omega}\times\mathbf{r}_j)^2+U(\mathbf{r}_j)]
\\
& +\frac{1}{2}\displaystyle\sum_{j\neq k}V_{jk},
\end{array} \label{ha}\end{equation}  where the notations are
standard, $U$ is the external potential, $V_{jk}\equiv
V(|\mathbf{r}_j-\mathbf{r}_k|)$ is the particle-particle interaction
and is rotationally invariant. For basic mechanics and
thermodynamics of a rotating body and the application to a Bose
system, we refer to the standard texts.~\cite{landau} But we draw
attention to the point that for each particle, the radius vector
$\mathbf{r}_j$, the canonical momentum $\mathbf{p}_j$, and the
angular momentum $\mathbf{l}_j=\mathbf{r}_j\times \mathbf{p}_j$ are
respectively the same in the laboratory frame and in the co-rotating
frame. It is for this reason that $H$ can be re-written as
$$H=H_{lab}-\bm{\omega}\cdot\sum_j \mathbf{l}_j, $$
where $$H_{lab}=\sum_{j} [\mathbf{p}_j^2/2m +U(\mathbf{r}_j)]
+(1/2)\sum_{j\neq k}V_{jk}$$ is the Hamiltonian in the laboratory
frame. This point is quite delicate in ODLRO study.~\cite{shi}

\begin{figure}[h]
\resizebox{5cm}{4cm}{\includegraphics[28pt,600pt][290pt,790pt]{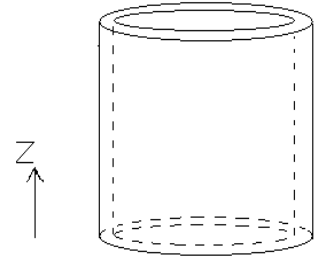}}
\caption{\label{pic} The cylindrical annular container as often
considered in literature and also here. The radius $R$ is much
larger than the thickness $d$. The rotation is along the axis of
the two concentric cylinders.}
\end{figure}

For simplicity, as usual, consider a thin cylindrical annular
container, with average radius $R$ and thickness $d \ll R$
(Fig.~\ref{pic}). The rotation $\bm{\omega}$ is, of course, along
the cylindrical axis ($z$ axis). Then the centrifugal potential
becomes a $\omega$-dependent constant (in the sense that it is
independent of the particle configuration), $-\frac{1}{2}M(\omega
R)^2$, where $M$ is the total mass of the particles.

It is probably useful to make a  synopsis here on the free energies
in the two reference frames and their relations with the rotational
inertial and the superfluid density. The free energy in the
co-rotating frame can be written as
\begin{equation} F= F_0-\frac{1}{2}I_{c}\omega^2 = a-\frac{1}{2}I\omega^2,
\label{f}
\end{equation}
where $a$ is a constant, $I_{c}=MR^2$ is the classical rotation
inertial,
$$F_0 \equiv a+\frac{1}{2}(I_{c}-I)\omega^2.$$
The total angular momentum is $\mathbf{L} =\langle \sum_j
\mathbf{l}_j\rangle$, hence
$$L_z=I\omega=-\frac{\partial F}{\partial \omega}.$$

In the laboratory frame, the free energy is $$F_{lab} =
F+\bm{\omega}\cdot\mathbf{L}=F+I\omega^2.$$ Therefore,
$$F_{lab} = F_{lab,0}+\frac{1}{2}I_{c}\omega^2= a+\frac{1}{2}I\omega^2,$$ where
$$F_{lab,0} \equiv a-\frac{1}{2}(I_{c}-I)\omega^2.$$
Consistently, one also has
$$L_z=I\omega=\frac{\partial F_{lab}}{\partial \omega}, $$
$$I =
-\frac{\partial^2 F}{\partial \omega^2}=\frac{\partial^2
F_{lab}}{\partial \omega^2}. $$

For a normal system, $I=I_{c}$, thus $F_0 = F_{lab,0} =a$. {\em If
$F_0$ or, equivalently, $F_{lab,0}$ depends on $\omega$, then the
system is a superfluid, with NCRI.} The superfluid fraction is
$$\frac{\rho_S}{\rho}=1-\frac{I}{I_c}=
\frac{1}{I_c}\frac{\partial^2 F_0}{\partial \omega^2}
=-\frac{1}{I_c}\frac{\partial^2 F_{lab,0}}{\partial \omega^2},
$$
where $\rho_S$ and $\rho$ are  the superfluid density and the
total fluid density,  respectively.

It should be noted that in equilibrium, it is $F$, not $F_{lab}$,
that is related to the partition function $Q$ as $F=-kT\ln Q$.

\section{A derivation of NCRI from ODLRO }

Now we make a general derivation that  if the system possesses
ODLRO, then $F_0$ in Eq.~(\ref{f}) depends on $\omega$. We use an
approach similar to Yang's treatment of superconductivity in a
magnetic field.~\cite{yang}

Using the cylindrical coordinates $(z,r,\theta)$ and considering
the geometry,  the Hamiltonian (\ref{ha}) can be simplified as
\begin{equation}
H= H_0-\frac{1}{2}M\omega^2 R^2, \label{h2}
\end{equation}
with
$$H_0 = \sum_j [\frac{(p_{\theta j}-m\omega R)^2}{2m}+
\frac{p_{zj}^2}{2m}] + \frac{1}{2}\sum_{j\neq k}
V_{jk}+NU(R),$$ 
where $p_{\theta j}=(1/R){\partial}/\partial \theta_j$. The radial
momentum $p_{rj}=\partial/\partial r_j$ is neglected  because  $d
\ll R$. An eigenfunction $\psi_{\alpha}$ of $H$ satisfies
$$H\psi_{\alpha} = E_{\alpha} \psi_{\alpha},$$ and the periodic
boundary condition, or SVBC,
\begin{equation}
\psi_{\alpha}(\theta_j+2\pi, \{\theta_{i\neq j}\}) =
\psi_{\alpha}(\theta_j,\{\theta_{i\neq j}\}) \label{pbc}
\end{equation} due to the cylindrical geometry.

Because  $-M\omega^2 R^2/2$ is a constant for a given
$\bm{\omega}$, we only need to consider $H_0$, whose
eigenfunctions are completely the same as those of $H $, i.e.
\begin{equation}
H_0\psi_{\alpha} = E_{\alpha}' \psi_{\alpha}, \label{eigen0}
\end{equation}
where
$E_{\alpha}'=E_{\alpha}+M\omega^2 R^2/2$.

By a ``gauge'' transformation
$$\psi_{\alpha} = \psi'_{\alpha} \exp (\frac{im\omega R\sum_j
\theta_j}{\hbar}),$$
$\psi'_{\alpha}$ satisfies
$$H_0'\psi_{\alpha}' = E_{\alpha}' \psi_{\alpha}',$$
where
$$H_0'=  \sum_j [\frac{p_{\theta j}^2}{2m}+ \frac{p_{zj}^2}{2m}] +
\frac{1}{2}\sum_{j\neq k} V_{jk}+NU(R).$$
The angular boundary condition becomes
\begin{equation}
\psi'_{\alpha}(\theta_j+2\pi, \{\theta_{i\neq j}\}) =
e^{-\frac{2\pi im\omega
R}{\hbar}}\psi'_{\alpha}(\theta_j,\{\theta_{i\neq j}\}).
\label{bc}
\end{equation}

Now consider the un-normalized density matrix
$$\rho_{dm}=e^{-\frac{H_0}{kT}}.$$
Because $H=H_0-I_c\omega^2/2$, $F_0$ in Eq.~(\ref{f}) is given by
$F_0=-kT\ln Q (H_0)$, where $Q(H_0)=Tr \rho_{dm}$. From
$\rho_{dm}$, by tracing over all but one particle, one obtains the
(un-normalized) one-particle reduced density matrix $\rho_1$.

The problem determined by $H_0$ together with SVBC is equivalent
to the description in terms of $H_0'$ together with
Eq.~(\ref{bc}). From the $\omega$-independence of $H_0'$ and the
boundary condition (\ref{bc}), one knows that
\begin{equation}
\langle \theta'+2\pi|\rho_1|\theta\rangle = \langle
\theta'|\rho_1|\theta-2\pi\rangle = e^{\frac{2\pi im\omega
R}{\hbar}}\langle \theta'|\rho_1|\theta\rangle. \label{r}
\end{equation}

We can now apply Yang's method to the current problem. Without
ODLRO, ${\rho}_1$ is vanishingly small except in the regions
around $\theta=\theta'\pm 2n\pi$, where $n=0, 1, \cdots$. As
indicated by Eq.~(\ref{r}), the values of $\rho_1$ in two
neighboring regions only differ by a phase factor $e^{\pm
\frac{2\pi im\omega R}{\hbar}}$.

With ODLRO, these regions with nonvanishing $\rho_1$ merge into
each other, and $\rho_1$ is nonvanishing everywhere. The above
phase relation remains.

Furthermore, with ODLRO, Eq.~(\ref{r})  implies that the
dependence of $\langle \mathbf{r}'|\rho_1|\mathbf{r}\rangle$ on
$\mathbf{r}-\mathbf{r}'$ must vary as $\omega$ varies.
Consequently, $Q(H_0)$ and thus $F_0$ also vary with $\omega$, as
$Q(H_0) =Tr_1\rho_1$ and $F_0= -kT\ln Q(H_0)$.  This proves that
ODLRO gives rise to superfluidity or NCRI.

\section{Quantization of angular momentum}

We now demonstrate the quantization of angular momentum as a
consequence of ODLRO, by employing the method of Bloch in
discussing superconducting persistent current,~\cite{bloch} and
also as a generalization of an argument by
Leggett.~\cite{leggett2} As said above, the angular momentum and
momentum are, respectively, the same in the laboratory frame and
in the co-rotating frame. But for convenience, here we use the
co-rotating frame.

Consider the one-particle reduced density matrix with $r$ and $z$
coordinates integrated over,
$$\langle \theta'|\rho_1|\theta\rangle = \int dr\int dz \langle
r,\theta',z|\rho_1|r,\theta,z\rangle, $$
and its
Fourier transformation
\begin{equation}
\langle \theta' |\rho_1|\theta\rangle = \frac{1}{2\pi} \sum_{l',l}
e^{\frac{i}{\hbar}(l\theta-l'\theta')}\langle l'|\rho_1|l\rangle,
\label{ft}
\end{equation} where $l$ and $l'$ represent angular
momenta. Its normalization is
$$\int \langle \theta|\rho_1|\theta\rangle d\theta = \sum
_{l}\langle l|\rho_1|l\rangle = N.$$
Conservation of angular momentum in the $z$ direction implies that
$\langle l'|\rho_1|l\rangle = 0$ for $l' \neq l$.

The total angular momentum, along the $z$ direction, for the
system under consideration can be given as
\begin{equation}
L_z = \sum_{l} l \langle l|\rho_1|l\rangle. \label{lz}
\end{equation}

In the Hamiltonian  in Eq.~(\ref{h2}), $p_{\theta j}$ can be
substituted as $l_{zj}/R$, where $l_{zj}$ is the $z$-component
angular momentum operator of the $j$-th particle. Thus $H$ depends
on single particle angular momentum operators through the kinetic
term
$$\sum_j (l_{zj}-m\omega
R^2)^2/(2mR^2).$$

Define $\tilde{l}=l-m\omega R^2$. In Eq.~(\ref{lz}), if the
summation can be replaced as an integral, then one can substitute
$l$ as $\tilde{l}+m\omega R^2$ and replace the integral over $l$
as that over $\tilde{l}$. Consequently one obtains $L_z=Nm\omega
R^2+L_z'(\omega)$, where $L_z'(\omega)$ is independent of $\omega$
and is thus equal to $L_z'(0)$, which must be vanishing. Therefore
$$L_z = N m\omega R^2,
$$
which is exactly the angular momentum of a classical object. But
is it legitimate to replace the summation over angular momentum
eigenvalues as an integral?

Let $\Delta \theta$ be the range of $|\theta'-\theta|$ in which
$\langle \theta' |\rho_1|\theta\rangle$ remains the same order of
magnitude as $\langle \theta |\rho_1|\theta\rangle$, while $\Delta
l$ be the half width of $\langle l |\rho_1|l\rangle$ around
$l=l_0$. Then because of Eq.~(\ref{ft}), we know
$$\Delta \theta \Delta l \approx \hbar.$$

In the absence of ODLRO, $\Delta\theta \ll 1$, thus
$$\Delta l \gg \hbar.$$
This allows the replacement of the summation over $l$ as an
integral, provided that $\langle l|\rho_1|l\rangle$ is smooth.

In contrast, the presence of ODLRO implies that $$\Delta\theta
\approx 1.$$ Therefore,
$$\Delta l \approx \hbar,$$
which is equal to the unit difference of angular momentum
eigenvalues. It is thus clear that if there is ODLRO, then one
cannot replace the summation as an integral.

For such a probability distribution caused by ODLRO,  $\langle l
|\rho_1|l\rangle \approx N_0$ for $l \approx l_0$, where  $N_0$ is
of the same order of magnitude of $N$, while $\langle l
|\rho_1|l\rangle$ for other values of $l$ are negligible. Thus the
total angular momentum is quantized as
$$L_z \approx N_0l_0,$$
with $l_0$  determined by minimizing the Hamiltonian. When $\omega$
is sufficiently small, $l_0=0$, i.e. the system exhibits
Hess-Fairbank effect. When $\omega$ is finite, $l_0$ is finite, but
$N_0l_0$ is less than $N m\omega R^2$.

\section{Re-derivation of Superfluid density in path integral Formalism}

The analyses on solid $^4$He in Refs.~\cite{ceperley,prokofev}
were based on an elegant path integral formulation of superfluid
density in a rotating annulus,~\cite{pollock} with the same
geometry as in our consideration above. It was derived by
neglecting the centrifugal potential. We believe that the
centrifugal potential cannot be neglected. As this formulation of
superfluid density is very important and widely used,  it may be
worthwhile to rederive it without neglecting the centrifugal
potential. It turns out that it nicely remains the same, although
the centrifugal potential is added to the free energy. But it
seems that this is known only after it is checked, so it is
reported here.

We re-write the Hamiltonian in the rotating frame, already given
in Eq.~(\ref{ha}), as
\begin{equation}
H = \sum_{j} \frac{(\mathbf{p}_j-m\mathbf{v})^2}{2m}
-\frac{1}{2}Nmv^2+ U+ V ,
\end{equation}
where,  to follow  Ref.~\cite{pollock},  the rotational velocity
$\omega R$ is denoted as $v$. The external potential  and the
interaction terms are schematically denoted as $U$ and $V$
respectively. $U$ is absent in Ref.~\cite{pollock}, but its addition
does not change the equations concerned. This Hamiltonian determines
the  density matrix $\rho_{dm}$ and the statistical distribution.

One obtains~\cite{pollock}
\begin{equation}
\frac{\rho_N}{\rho}Nm\mathbf{v} =
\frac{Tr(\mathbf{P}\rho_{dm})}{Tr(\rho_{dm})}, \label{n}
\end{equation}
where   $\rho_N$  is the normal fluid density, $\mathbf{P}=\sum_j
\mathbf{p}_j$ is the total momentum. This identity is obtained by
considering the momentum in the laboratory frame, as $\mathbf{v}$
is the container velocity in the laboratory frame. Again, note
that the canonical momentum is the same in the laboratory and in
the co-rotating frames, while it reduces to the kinematic momentum
in the laboratory frame.

Because
$$\mathbf{P} = -\frac{\partial H}{\partial \mathbf{v}},$$
Eq.~(\ref{n}) can be re-written as
\begin{equation}
\frac{\rho_N}{\rho}Nm\mathbf{v} = -\frac{\partial F}{\partial
\mathbf{v}},
\end{equation}
where $ F=-kT\ln[Tr(\rho_{dm})]$ is the free energy in the
co-rotating frame. Therefore the superfluid fraction is
$$\frac{\rho_S}{\rho} = 1+\frac{\partial (\frac{F}{N})}{\partial
(\frac{1}{2}mv^2)}.$$ Thus the free-energy change due to the
rotation of the container, up to the order of $v^2$, is
\begin{equation}
\frac{\Delta F}{N} = \frac{mv^2}{2}(\frac{\rho_S}{\rho}-1),
\label{rhos} \end{equation} from which it can be confirmed that
the centrifugal potential $mv^2/2$ indeed cannot be ignored, since
it is no less than the other term  $(\rho_S/\rho)mv^2/2$.

In the path integral calculation,
$$ e^{-\beta\Delta F} =
\frac{\int\rho_{dm}(\mathbf{X},\mathbf{X};\beta;\mathbf{v})d\mathbf{X}}
{\int\rho_{dm}(\mathbf{X},\mathbf{X};\beta;\mathbf{v}=0)d\mathbf{X}},$$
where $\mathbf{X}$ represents the configuration of the particles.
The ``gauge term'' $-m\mathbf{v}$ in the kinetic energy term can
be transformed away, by adding, in the density matrix elements, a
phase factor in winding the periodic system, like in
Eqs.~(\ref{bc}) and (\ref{r}). Consequently, one can replace
$\rho_{dm}(\mathbf{X},\mathbf{X};\beta;\mathbf{v})$ as the density
matrix $\tilde{\rho}_{dm}(\mathbf{X},\mathbf{X};\beta;\mathbf{v})$
corresponding to the Hamiltonian without the ``gauge term'', while
multiplying it by a phase factor due to the total paths
$\mathbf{W}L$ of the $N$ particles winding around the system,
where  $\mathbf{W}$ is the winding number.~\cite{pollock}

$\tilde{\rho}_{dm}=\exp (-\beta \tilde{H})$, where $\tilde{H} = H
(\mathbf{v}=0)-Nmv^2/2$. $H (\mathbf{v}=0)$ is just $H_0'$ in Sec.
III. We obtain
$$
e^{-\beta\Delta F} = \langle
e^{i\frac{m}{\hbar}\mathbf{v}\cdot\mathbf{W}L}
e^{\beta\frac{1}{2}Nmv^2}\rangle,$$ where the average that of the
density matrix with $\mathbf{v}=0$. Consequently, up to the order of
$v^2$, we have
$$\Delta F = N\frac{mv^2}{2}(\frac{m\langle W^2\rangle
L^2}{3\beta\hbar^2 N}-1),$$ which, together with Eq.~(\ref{rhos}),
yields
$$\frac{\rho_S}{\rho} =\frac{m\langle W^2\rangle
L^2}{3\beta\hbar^2 N},$$ which is the same as that given in
Ref.~\cite{pollock} They remain the same even if $v$ is not a
small quantity, for the reason is that $v$ is independent of the
particle configuration.

\section{No ODLRO in Nosanow-Jastrow wavefunctions}

Now we turn our attention to solid $^4$He. For a commensurate
solid at rest,  each atom occupies a lattice site. Because of
quantum mechanical zero-point motion, which is large in solid
helium, around the neighborhood of each lattice site, there is a
finite region in which the wavefunction is nonvanishing.  With the
exchange effect put aside first, the wavefunction  is localized
around each lattice site, i.e. it decays from the maximum at the
lattice site. Let's denote the wavepacket of  atom $i$ as
$w(\mathbf{r}_i-\mathbf{Q}_i)$, where $\mathbf{r}_i$ is the actual
position of the atom, $\mathbf{Q}_i$ represents a lattice site
fixed in the solid. The Hartree approximation of the wavefunction
of the solid helium is the product of these single-atom
wavefunctions, i.e.,
\begin{equation}
\Phi_{H}=\prod_{i=1}^{N}w(\mathbf{r}_i-\mathbf{Q}_i),
\end{equation} which was indeed used in the earliest
(unsatisfactory) variational calculations of solid
helium.~\cite{nosanow1} Later works, starting by
Nosanow,~\cite{nosanow2} took into account the two-particle
short-range correlation by multiplying the Hartree wavefunction by
the Jastrow factor.

To account for the exchange effect due to overlap between
neighboring single-particle wavepackets, one also needs to consider
the wavefunction symmetrized over all the atoms; the detailed nature
of the exchange effect is then determined by the Hamiltonian.  With
symmetrization, the Hartree approximation is improved to
Hartree-Fock approximation,
\begin{equation}
\Phi_{HF} = \frac{1}{\sqrt{N!}} \sum_{P} P
\prod_{i=1}^{N}w(\mathbf{r}_i-\mathbf{Q}_i), \label{hf}
\end{equation}
where $P$ represents $N!$ permutations of the $N$ lattice sites
$\{\mathbf{Q}_i\}$. The symmetrization can be made on either the
particle positions $\{\mathbf{r}_i\}$ or the lattice sites
$\{\mathbf{Q}_i\}$. We choose the latter for easier manipulation
below.

The symmetrized Nosanow-Jastrow wavefunction is
\begin{equation}
\Phi_{SNJ}= K \sum_{P} P
\prod_{i=1}^{N}w(\mathbf{r}_i-\mathbf{Q}_i) \prod_k \prod_{j<k}
f_{jk},\label{nj}
\end{equation}
where $K$ is the normalization constant,
$$f_{jk} \equiv f(-u(|\mathbf{r}_j-\mathbf{r}_k|))$$ is the Jastrow
(or, to be historically precise, Bijl-Dingle-Jastrow) function.
$f(-u(r))$ attains a maximum larger than $1$ at a certain distance
$r_0$,  and it is constrained to be $f \rightarrow 0$ as
$r\rightarrow 0$, and $f \rightarrow 1$ as $r\rightarrow \infty$
or $r>\sigma$ where $\sigma$ is a parameter. Note that $ \prod_k
\prod_{j<k} f_{jk}$ is automatically symmetric for all particles.

Our consideration is about a thin cylindrical bulk, $\mathbf{r}_i=
(R, R\theta_i, z_i)$. Especially,  the periodic boundary condition
in coordinate $\theta $ should be taken into account in an
essential way. It implies that the Wannier-like function $w$ must
be of the form~\cite{kohn2}
\begin{equation}
w(\mathbf{r}-\mathbf{Q})= A \sum_{\gamma=-\infty}^{\infty}\bar{w}
(\mathbf{r}-\mathbf{Q}-\gamma\mathbf{G}), \label{w}
\end{equation}
where $A$ is the normalization constant, $\mathbf{G}=2\pi
R\hat{\theta}$ represents the circumference, $\gamma$ represents
integers, $\bar{w}$ is the (real) Wannier-like function for the
infinite interval. Each $\bar{w}$ extends over a finite range,
much smaller than the system size, but finite overlap is allowed.
$\pm \infty$ in the summation can be understood as two bounds
which can be arbitrarily large. Thus
\begin{equation}
\bar{w}(\mathbf{r}) \bar{w}(\mathbf{r}-\mathbf{S}) \approx
\bar{w}^2 (\mathbf{r}) \exp (-|\mathbf{S}|/c), \label{m}
\end{equation}
where $\mathbf{S}$ is an arbitrary vector, $c$ is a length scale
less than the lattice constant. Consequently, the normalization
constant $A$ in Eq.~(\ref{w}) is  $A \approx
(\sum_{\gamma,\gamma'} \exp (-|\gamma-\gamma'|G/c))^{-1/2}$.

Moreover, it can be found that
\begin{equation}
w(\mathbf{r}) w(\mathbf{r}-\mathbf{S}) \leq \bar{w}^2 (\mathbf{r})
\exp (-|\bar{S}_{\theta}|/c), \label{mw}
\end{equation}
where $\bar{S}_{\theta}$ is the  $\theta$ component of
$\mathbf{S}$ modulo $\pm G$ such that $|\bar{S}_{\theta}| \leq
G/2$, i.e., $|\bar{S}_{\theta}|$ is the shortest
$\theta$-component of the distance between the two physical points
represented by $\mathbf{r}$ and $\mathbf{r}-\mathbf{S}$.

We now set out to  show that there is no ODLRO or BEC in $\Phi_H$
or $\Phi_{HF}$ or $\Phi_{SNJ}$, by examining  the one-particle
reduced density matrix
$$\rho_{1}(\mathbf{r},\mathbf{r}')=N\int d\mathbf{r}_2\cdots
d\mathbf{r}_N \Phi(\mathbf{r},\mathbf{r}_2,\cdots,\mathbf{r}_N)
\Phi(\mathbf{r}',\mathbf{r}_2,\cdots,\mathbf{r}_N)$$ for the
ground state wavefunction $\Phi$ of the form of  $\Phi_H$ or
$\Phi_{HF}$ or $\Phi_{SNJ}$.

Though trivial, it is instructive to first consider  $\Phi_H$. It
is straightforward to integrate out
$\mathbf{r}_2,\cdots,\mathbf{r}_N$, and obtain
$\rho_1(\mathbf{r},\mathbf{r}')=Nw(\mathbf{r}-\mathbf{Q}_1)
w(\mathbf{r}'-\mathbf{Q}_1)$, for which Eq.~(\ref{mw}) directly
leads to
\begin{equation}
\rho_1(\mathbf{r},\mathbf{r}') \leq N \bar{w}^2 (\mathbf{r}) \exp
(-|\overline{x-x'}|/c), \label{mh}
\end{equation}
where $x =R\theta$ denotes the $\theta$-component of $\mathbf{r}$.
Of course, $\bar{w}^2 (\mathbf{r}) \leq 1$. Thus
$\rho_1(\mathbf{r},\mathbf{r}') \rightarrow 0$ as
$|\overline{x-x'}|$ approaches the system size, i.e. there is no
ODLRO or BEC in $\Phi_H$.

Now consider the Hartree-Fock wavefunction $\Phi_{HF}$. In the
expansion of $\rho_1$, suppose the lattice sites in the first
$\Phi$ are denoted as $\{\mathbf{Q}_i\}$ while those in the second
$\Phi$ are denoted as $\{\mathbf{Q}_i'\}$. The exponential decay
of the overlap between single-particle wavefunctions,
Eq.~(\ref{mw}), implies that among the $(N!)^2$ terms in the
expansion of $\rho_1$, one can neglect each term in which
$\mathbf{Q}_i \neq \mathbf{Q}_i'$ for at least one of
$i=2,\cdots,N$.  Consequently, there are only $N!$ remaining
terms, in each of which $\mathbf{Q}_i = \mathbf{Q}_i'$ for
$i=1,\cdots,N$, then $\mathbf{r}_2,\cdots,\mathbf{r}_N$ are
subsequently all integrated out. This $N!$ is cancelled by the
$N!$ in the normalization constant. Hence, for large
$|\overline{x-x'}|$, $\rho_1(\mathbf{r},\mathbf{r}')$ for
$\Phi_{HF}$ behaves in the same way as for the Hartree
wavefunction, given in Eq.~(\ref{mh}). This proves there is no
ODLRO or BEC in $\Phi_{HF}$ either.

The argument can be extended to  symmetrized Nosanow-Jastrow
wavefunction $\Phi_{SNJ}$, which  can be re-written as
\begin{equation}
\Phi_{SNJ}= K \sum_{P}P
\prod_{i=1}^{N}[w(\mathbf{r}_i-\mathbf{Q}_i) \prod_{j<i} f_{ji}],
\label{nj2}
\end{equation}
where $P$ represents the permutation of the $N$ lattice sites
$\{\mathbf{Q}_i\}$. $\prod_{j<i} f_{ji}$ is a function of
$\mathbf{r}_1,\cdots, \mathbf{r}_{i}$, and reduces to $1$ for
$i=1$. For each term in the expansion of $\rho_1$, consider
$w(\mathbf{r}_i-\mathbf{Q}_i)
w(\mathbf{r}_i-\mathbf{Q}_i')(\prod_{j<i} f_{ji})^2 \leq \bar{w}^2
(\mathbf{r}) \exp
(-|\overline{Q_{i\theta}-Q_{i\theta}'}|/c)(\prod_{j<i} f_{ji})^2
$, where $Q_{i\theta}$ is the $\theta$ component of
$\mathbf{Q}_{i}$. It can be seen that the short-range Jastrow
factor does not change the nature of long-range exponential decay.
Therefore, the cross terms, in which $\mathbf{Q}_{i} \neq
\mathbf{Q}_{i}'$ for at least one of $i=1,\cdots,N$, exponentially
decay, and are negligible in comparison with the remaining terms.
Consequently,
\begin{widetext}
\begin{eqnarray}
\rho_1(\mathbf{r},\mathbf{r}')& \approx&
\frac{Nw(\mathbf{r}_1-\mathbf{Q}_1)w(\mathbf{r}_1'-\mathbf{Q}_1)
\prod_{i>1} \int w^2 (\mathbf{r}_i-\mathbf{Q}_i) (\prod_{1<j<i}
f_{ji})^2 f_{1i}f_{1i}' d\mathbf{r}_i} {\prod_{i} \int w^2
(\mathbf{r}_i-\mathbf{Q}_i) (\prod_{j<i} f_{ji})^2 d\mathbf{r}_i}\\
&\leq & N\bar{w}^2(\mathbf{r}_1)e^{-|\overline{x-x'}|/c} \frac{
\prod_{i>1} \int w^2 (\mathbf{r}_i-\mathbf{Q}_i) (\prod_{1<j<i}
f_{ji})^2 f_{1i}f_{1i}' d\mathbf{r}_i} {\prod_{i} \int w^2
(\mathbf{r}_i-\mathbf{Q}_i)(\prod_{j<i} f_{ji})^2 d\mathbf{r}_i},
\label{fract}
\end{eqnarray}
\end{widetext}
where $f_{1i}'\equiv f(-u(|\mathbf{r}_1'-\mathbf{r}_i|))$. The
fraction factor in (\ref{fract}) must be bounded by a finite
number. Clearly, $\rho_1(\mathbf{r},\mathbf{r}')$ tends to
exponentially vanish as $|\overline{x-x'}|$ approaches the system
size. Thus there is no ODLRO or BEC in $\Phi_{SNJ}$ either. It can
be  seen that our argument is not disrupted by the thermodynamic
limit $N\rightarrow \infty$.

Furthermore, the argument can be straightforwardly generalized to
a finite temperature, in which each energy eigenfunction is of the
form of $\Phi_H$ or $\Phi_{HF}$ or $\Phi_{SNJ}$. The
finite-temperature density matrix is the thermal average of the
density matrices of the eigenfunctions. For an infinite sample,
$w$ would simply be $\bar{w}$, the conclusion of no ODLRO can
still be obtained, in a simpler way.

Therefore, although there is ODLRO or BEC in the Jastrow
wavefunction alone, which describe liquid helium,~\cite{mcmillan}
they are dominated by the localized one-particle wavefunctions. This
is a difference between liquid and solid. The argument extends that
of Penrose an Onsager about no BEC in a solid~\cite{penrose} to the
case with zero-point motion, exchange effect, as well as short-range
correlation.

\section{No Supersolidity in Nosanow-Jastrow wavefunctions}

As ODLRO is a sufficient condition of NCRI, it is not redundant to
demonstrate that there is no NCRI  in $\Phi_H$ or $\Phi_{HF}$ or
$\Phi_{SNJ}$, as we now explicitly do in the following. We adapt the
method of Kohn used in discussing electronic insulating
state.~\cite{kohn2}

Recall that the eigenfunctions and energy spectrum is determined by
$H_0$, as in Eq.~(\ref{eigen0}). The idea is the following. For
every eigenfunction $\Psi_{\alpha}(\omega=0)$ of $H_0(\omega=0)$,
where $\alpha$ is the index for different eigenfunctions, be it of
the form of $\Phi_H$ or $\Phi_{HF}$ or $\Phi_{SNJ}$, we show that
there is a corresponding eigenfunction $\Psi_{\alpha}(\omega \neq
0)$ of $H_0(\omega \neq 0)$, and that its eigenvalue remains the
same as that of $H_0(\omega =0 )$ for $\Psi_{\alpha}(\omega = 0)$.

$H_0(\omega \neq 0)$ is simply related to $H_0(\omega=0)$ by a
``gauge'' transformation, but one should be cautioned by the
requirement of the SVBC,~\cite{leggett0,kohn2} Eq.~(\ref{pbc}). In
an infinite interval, for a localized single-particle
eigenfunction $\bar{w}(\mathbf{r})$ of a single-particle
Hamiltonian, the correct eigenfunction  wavefunction for $\omega
\neq 0$ is
$$\bar{w}'(\omega;\mathbf{r})=
\bar{w}(\mathbf{r}) \exp (\frac{im\omega x}{\hbar}),$$ where, as
above, $x=R\theta$.

Therefore, for a many-particle eigenfunction
$\Psi_{\alpha}(\omega=0)$ of $H_0(\omega=0)$, given by $\Phi_H$ or
$\Phi_{HF}$ or $\Phi_{SNJ}$, one may construct the corresponding
eigenfunction $\Phi_{\alpha}(\omega \neq 0)$ of $H_0(\omega \neq
0)$ in a similar way, by replacing every single-particle
$\bar{w}(\mathbf{r})$ as $\bar{w}'(\mathbf{r})$. The presence of
Jastrow factor does not affect this.

On the other hand, by using Eq.~(\ref{w}),
$\Psi_{\alpha}(\omega=0)$ can be written as
$$\Psi_{\alpha} (\omega=0) = A^N\sum_{\Gamma=-\infty}^{\infty}
\bar{\Phi}_{\Gamma}(\{\mathbf{r}_i\}),$$
where $\bar{\Phi}_{\Gamma}$ is obtained from $\bar{\Phi}$ by
shifting the centers of the single-particle wavepackets $\bar{w}$
from $\{\mathbf{Q}_i\}$ to $\{\mathbf{Q}_i+\gamma_i\mathbf{G}\}$,
with $\sum_i \gamma_i = \Gamma$; here $\bar{\Phi}$ is of the form
of $\bar{\Phi}_{H}= \prod_{i=1}^{N}
\bar{w}(\mathbf{r}_i-\mathbf{Q}_i)$, or $\bar{\Phi}_{HF} =
\frac{1}{\sqrt{N!}} \sum_{P} P
\prod_{i=1}^{N}\bar{w}(\mathbf{r}_i-\mathbf{Q}_i),$ or
$\bar{\Phi}_{SNJ}= K \sum_{P} P
\prod_{i=1}^{N}\bar{w}(\mathbf{r}_i-\mathbf{Q}_i) \prod_k
\prod_{j<k} f_{jk}.$

Following the argument in Ref.~\cite{kohn2}, using the exponential
decay of the overlap as given in Eq.~(\ref{m}),  and very similar
to the argument in last section, it can be shown that for $\Gamma
\neq \Gamma'$ and arbitrary $\alpha$ and $\alpha'$,
$\Phi_{\alpha,\Gamma}$ and $\Phi_{\alpha',\Gamma'}$ have
exponentially vanishing overlap and give vanishing matrix element
for an arbitrary one-particle position operator.

Consequently, it can be found that the corresponding eigenfunction
of $H_0(\omega)$, satisfying the SVBC, is
\begin{equation}
\Psi_{\alpha} (\omega) = \sum_{\Gamma =-\infty}^{\infty}
\Phi_{\alpha,\Gamma}(\{\mathbf{r}_i\}) \exp[\frac{im\omega
R}{\hbar}(\sum_j \theta_j-2\pi \Gamma)].
\end{equation}
Because of exponentially
vanishing overlap between $\Phi_{\alpha,\Gamma}$ with different
values of $\Gamma$, it is clear that
$$H_0(\omega)\Psi_{\alpha} (\omega) =
E_{\alpha}(\omega)\Psi_{\alpha}(\omega).$$ with
$$E_{\alpha}(\omega) = E_{\alpha}(\omega=0).$$

It is thus proved that every eigenvalue $E_{\alpha}(\omega)$ of
$H_0(\omega)$ is independent of $\omega$. Interestingly,  the
argument has gone through even in presence of the Jastrow factors.

In fact, the explicit construction of the wavefunction here
confirms the principle, established by Leggett,~\cite{leggett0}
that the system is non-superfluid if for the wavefunction of the
rotating system, the SVBC can still be satisfied without causing
the energy to be increased by the rotation. Leggett already
applied this principle to the Hartree-Fock wavefunction.

Therefore, for a commensurate quantum solid described by Hartree
or Hartree-Fock or Nosanow-Jastrow wavefunction, even though the
exchange effect, large zero-point motion and short-range
correlation are taken into account, the free energy is of the form
of Eq.~(\ref{f}), with $F_0$ independent of $\omega$. This
indicates that it cannot a supersolid.

In our argument, the localized single-particle wavefunctions play
a crucial role. Obviously,  the situation would be different when
there exist vacancies or interstitials or both, which makes the
wavefunctions extended. The recent experimental result of Kim and
Chan poses a significant challenge. The difficulty might be
resolved if an extended factor is found in the actual
wavefunction.

\section{Summary}

To summarize, we have offered some analytic arguments concerning
the existence or non-existence of superfluidity or supersolidity
behavior. This work might be useful for further investigations on
the cause of supersolidity. It might be helpful in supplementing
the understanding of the relevant classic literature, and in
clarifying which and which specific features are counterparts
between superfluidity and superconductivity.

Our argument seems to suggest that ODLRO is indeed generically
sufficient for superfluidity even in a finite temperature, a
question which seems to have remained not entirely resolved
previously.

Our discussions start with a synopsis, in Sec. II, on the
Hamiltonians and the free energies in the co-rotating and the
laboratory reference frames, as well as their relations with
rotational inertial and superfluidity density.

In Secs. III and IV, from the presence of ODLRO, we make a general
derivation of the most basic manifestation of superfluidity,
namely the Hess-Fairbank effect or NCRI, i.e., the refusal of the
Bose system to follow the rotation of the container, by using a
method of Yang in treating superconductivity in a magnetic field.
We also derive the quantization of angular momentum as a
consequence of ODLRO, by borrowing a method of Bloch in studying
superconducting persistent current. In Sec. V, we rederive the
path integral formulation of the superfluid density without
neglecting the centrifugal potential.

In Secs. VI and VII, we consider the variational wavefunctions
which have been used in solid helium calculations, namely, the
Hartree, the Hartree-Fock  and the symmetrized Nosanow-Jastrow
wavefunctions. The non-superfluidity in the Hartree-Fock
wavefunctions  was already noted by Leggett from its
disconnectivity.~\cite{leggett0} We show that there is no ODLRO in
these trial  wavefunctions, for both an infinite sample and that
confined in a cylindrical annulus. Moreover, by extending a method
originally due to Kohn in discussing electronic insulating states,
we explicitly demonstrate that there is no NCRI behavior in a
commensurate quantum solid described by  those trial
wavefunctions, even if there exist large zero-point motion, finite
overlap between wavepackets and exchange effect. In this argument,
the  constraint of SVBC in the angular direction is carefully
taken into account. The explicit construction of the wavefunction
under the rotation is consistent with the early arguments of
Leggett in terms of the connectivity
properties.~\cite{leggett0,leggett3}

\acknowledgements

I am grateful to Professor Sir Tony Leggett for useful comments,
as well as hospitality at UIUC in the academic year 2003-2004. I
am grateful to Prof. Chen-Ning Yang, Prof. Hwa-Tung Nieh and other
faculty members for current hospitality at CASTU. I thank Zheng-Yu
Weng, Yong-Shi Wu and Hui Zhai for related conversations.

\newpage

\begin{large}
{\bf Erratum}
\end{large}

[published as Phys. Rev. B {\bf 74}, 029901 (E) (2006)]

\vspace{1cm}

In Secs.~VI and VII,  the treatment of symmetrization was
inadequate. The problem is fixed below. Our condition is more
relaxed: the conclusion holds as far as the overlap between two
atomic Wannier functions decays exponentially, with the decay width
$c$ much less than the system size, rather than the lattice
constant. That is, $\bar{w}(\mathbf{r})
\bar{w}(\mathbf{r}-\mathbf{S}) \approx \bar{w}^2 (\mathbf{r}) \exp
[-|\mathbf{S}|/c]$,  for the Wannier function  $\bar{w}$ in an
infinite interval, or $w(\mathbf{r}) w(\mathbf{r}-\mathbf{S}) \leq
\bar{w}^2 (\mathbf{r}) \exp [-|\Theta(\mathbf{S})|/c]$ for the
Wannier function $w$ in the annular geometry, where
$\Theta(\mathbf{S})$ denotes the shortest $\theta$-component of
$\mathbf{S}$. As in the original paper, we discuss the annular
geometry; the case of infinite interval is simpler.

Consider the Hatree-Fock wavefunction
%
$\Phi_{HF} = D  \sum_{P}
\prod_{i=1}^{N}w(\mathbf{r}_i-\mathbf{Q}_{P_i}),$
where $P$ represents $N!$ permutations of the $N$ lattice sites
$\{\mathbf{Q}_i\}$,  $\{P_i\}$ are the indices for the $N$ lattice
sites in permutation $P$.  It can be found that
$D=1/N!\sqrt{\Delta}$, $\Delta = \sum_P  \prod_i O_{iP_i}$, where
$O_{iP_i}=\int d \mathbf{r}_i w(\mathbf{r}_i-\mathbf{Q}_i)
w(\mathbf{r}_i-\mathbf{Q}_{P_i})$. The one-particle density matrix
is thus found to be \begin{widetext}
\begin{eqnarray} \rho_{1}(\mathbf{r},\mathbf{r}') & =  &
\frac{N}{N!\Delta} \sum_{P,P'}
w(\mathbf{r}-\mathbf{Q}_{P_1})w(\mathbf{r}'-\mathbf{Q}_{P'_1})
\prod_{i\neq 1} \int d\mathbf{r}_i
w(\mathbf{r}_i-\mathbf{Q}_{P_i})w(\mathbf{r}_i-\mathbf{Q}_{P'_i} \nonumber) \\
& \leq &\frac{N}{N!\Delta} \sum_{P,P'}
w(\mathbf{r}-\mathbf{Q}_{P_1})w(\mathbf{r}'-\mathbf{Q}_{P'_1})
\prod_{i\neq 1} \exp [-|\Theta(\mathbf{Q}_{P_i}-\mathbf{Q}_{P'_i})|/c] \nonumber \\
&\leq & \frac{N}{N!\Delta} \sum_{P,P'}
\bar{w}^2(\mathbf{r}-\mathbf{Q}_{P_1})\exp
(-|\Theta(\mathbf{r}-\mathbf{r}'-\mathbf{Q}_{P_1}+\mathbf{Q}_{P'_1})|/c)\exp
[-|\Theta(\mathbf{Q}_{P_1}-\mathbf{Q}_{P'_1})|/c]  \nonumber \\
&\leq& \frac{N\cdot N!}{\Delta}\exp
[-|\Theta(\mathbf{r}-\mathbf{r}')|/c]. \nonumber \end{eqnarray}
\end{widetext}
In the derivation, the relation $\sum_i \mathbf{Q}_{P_i} = \sum_i
\mathbf{Q}_{P'_i}$ has been used. Therefore,
$\rho_{1}(\mathbf{r},\mathbf{r}') \rightarrow 0$ as
$|\Theta(\mathbf{r}-\mathbf{r}')|$ approaches the system size. The
argument is valid in the thermodynamic limit, as $\Delta$ is of the
same order of magnitude of $N!$.

Now consider the symmetrized Nosanow-Jastrow wavefunction
$\Phi_{SNJ}= K_N \sum_{P}  \prod_{i=1}^{N}
w(\mathbf{r}_i-\mathbf{Q}_{P_i}) \prod_{j<k} f_{jk}$, where $f_{jk}
\equiv f(|\mathbf{r}_j-\mathbf{r}_k|)$ is the Jastrow function,
$K_N^{-2}=\sum_{P,P'}\prod_l \int d\mathbf{r}_l \prod_i
w(\mathbf{r}_i-\mathbf{Q}_{P_i})w(\mathbf{r}_i-\mathbf{Q}_{P'_i})(\prod_{j<k}
f_{jk})^2$. In a way similar to the above derivation for
$\Phi_{HF}$, it is found that
$\rho_{1}(\mathbf{r},\mathbf{r}')  \leq   N K_N^2 \sum_{P,P'}
w(\mathbf{r}-\mathbf{Q}_{P_1})w(\mathbf{r}'-\mathbf{Q}_{P'_1})\exp
[-|\Theta(\mathbf{Q}_{P_1}-\mathbf{Q}_{P'_1})|/c] \prod_{l \neq 1}
\int d\mathbf{r}_l \prod_{i\neq 1}
f(|\mathbf{r}-\mathbf{r}_i|)f(|\mathbf{r}'-\mathbf{r}_i|)
w^2(\mathbf{r}_i-\mathbf{Q}_{P_i})(\prod_{1<j<k} f_{jk})^2  \leq N
\exp [-|\Theta(\mathbf{r}-\mathbf{r}')|/c]F^2 K_N^2\sum_{P,P'}
\prod_{i\neq 1} \int d\mathbf{r}_i
w^2(\mathbf{r}_i-\mathbf{Q}_{P_i})(\prod_{1<j<k} f_{jk})^2$, where
$F$ is the maximum value of the Jastrow function. It can be seen
that $\rho_{1}(\mathbf{r},\mathbf{r}') \rightarrow 0$ as
$|\Theta(\mathbf{r}-\mathbf{r}')|$ approaches the system size, and
that the argument is valid in the thermodynamic limit.

Thus it is proved that there is no ODLRO in $\Phi_{HF}$ or
$\Phi_{SNJ}$ under the condition stated above.  The symmetrization
should be dealt with in a similar way in Sec.~VII, where it is shown
that $\Phi_{\alpha,\Gamma}$ and $\Phi_{\alpha',\Gamma'}$, with
$\Gamma \neq \Gamma'$, have exponentially vanishing overlap and give
vanishing matrix element for an arbitrary one-particle position
operator, which leads to the conclusion of no supersolidity in these
wavefunctions.

As noted in Ref.~1, it had been shown previously that there is no
ODLRO in Hatree-Fock wavefunctions under the condition that the sum
of the overlap integrals between any site and its neighbors is less
than unity~\cite{schwartz}.

I thank Xin-Cheng Xie for questioning the original treatment of the
symmetrization.

\end{document}